%% file: TimeDistortionModeling.tex
\documentclass{sig-alternate-without-copyright}

\usepackage{url}
\usepackage{pgfplots}
\usepackage{enumitem}

\begin{document}

\title{Leveraging Time Distortion for seamless Navigation into Data Space-Time Continuum}

\numberofauthors{2} 
\author{
\alignauthor Thomas Hartmann and Francois Fouquet and Yves Le Traon\\
       \affaddr{SnT - University of Luxembourg}\\
       \email{first.last@uni.lu}
\alignauthor Brice Morin\\
       \affaddr{SINTEF ICT Norway}\\
       \email{brice.morin@sintef.no}\\    
}

\maketitle
\begin{abstract}
Intelligent software systems continuously analyze their surrounding environment and accordingly adapt their internal state.
Depending on the criticality index of the situation, the system should dynamically focus or widen its analysis and reasoning scope.
A naive ---{\em why have less when you can have more}--- approach would consist in systematically sampling the context at a very high rate and triggering the reasoning process regularly.  
This reasoning process would then need to mine a huge amount of data, extract a relevant view, and finally analyze this adequate view.
This overall process would require some heavy resources and/or be time-consuming, conflicting with the (near) real-time response time requirements of intelligent systems.
We claim that a continuous and more flexible navigation into context models, in space and in time, can significantly improve reasoning processes. 
This paper thus introduces a novel modeling approach together with a navigation concept, which consider time and locality as first-class properties crosscutting any model element, and enable the seamless navigation of models in this space-time continuum.
In particular, we leverage a time-relative navigation (inspired by the space-time and distortion theory~\cite{hawking1973large}) able to efficiently empower continuous reasoning processes.  
We integrate our approach into an open-source modeling framework and evaluate it on a smart grid reasoning engine for electric load prediction.
We demonstrate that reasoners leveraging this distorted space-time continuum outperform the full sampling approach, and is compatible with most of (near) real-time requirements.

\end{abstract}
%

\input{introduction}
\input{background}
\input{contribution}
\input{evaluation}
\input{relatedwork}
\input{conclusion}

\bibliographystyle{abbrv}
\bibliography{sigproc}  

\end{document}

%% file: introduction.tex


\section{Introduction}

An intelligent software system needs to analyze both its surrounding environment and its internal state, which together we refer to as the context of the system, in order to continuously adapt itself to varying conditions.
Therefore, building an appropriate context model, which reflects the current context of the system is of key importance.
This task, is however not trivial, as different reasoning processes need to leverage models with different granularities: some fine-grained models relying on local and instantaneous data to handle critical situations, while some others relying on broader and coarser-grained models to analyze trends and achieve a large-scale consensus.   

Let us make an analogy with a chess player. A chess player has deep knowledge of all the famous games played in the past, which he can utilize when he is playing a game. However, for example in a check situation, he is forced to focus his reasoning around his king.
When the time is running out the player can hardly leverage his complete historical background and rather focuses his attention on the current situation.
In such situations the player focuses his attention to a restricted number of moves and pawns in a game.
From the same context (a person playing chess), different views are thus extracted according to the current situation and the criticality index of the context, in order to conduct appropriate reasoning.

\begin{figure}[ht]
\centering
\begin{tikzpicture}[scale=0.65]
\begin{axis}[
xlabel=model space,
xlabel style={rotate=-8},
ylabel=model space,
ylabel style={rotate=39},
zlabel=time space, clip=false, xtick=\empty, ytick=\empty, 
ztick={1,2},
zticklabels={t$_{i}$,t$_{i+1}$},
]

\addplot3[surf, samples=10, color=white]
{1};
\addplot3+[mesh, scatter, samples=10, mark=*, color=black]
coordinates {
	(-4,0,1) (-3,-4,1) (4,-4,1) (0,0,1) (1.5,3,1) (-3,3,1) (0,0,1) (2.5,0,1)
};

\addplot3[surf,samples=10, color=white]
{2};
\addplot3+[mesh, scatter, samples=10, mark=*, color=black]
coordinates {
	(-4.5,3.5,2) (-4.5,-4,2) (4,-4,2) (0,0,2) (2,4,2) (-3,1,2) (0,0,2) (4,4,2)
};
\end{axis}
\end{tikzpicture}
\vspace{-1.0em}
\caption{Linear Sampled Model}
\label{linear_time}
\end{figure}
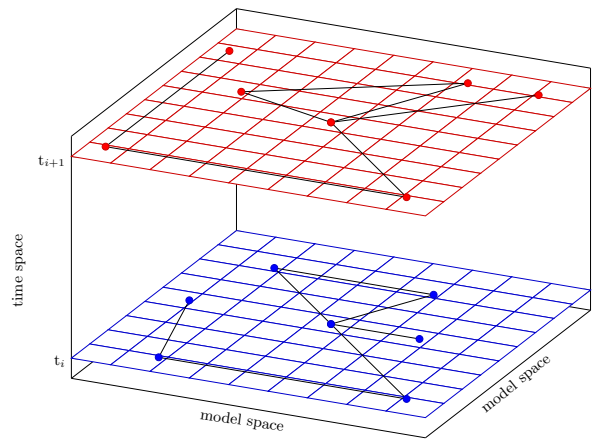

Back to the software engineering domain, let us take a smart grid reasoning engine as an example. Due to changes in the production/consumption chain over time, or due to the sporadic availability of natural resources (heavy rain or heavy wind), the properties of the smart grid must be continuously monitored and adapted to regulate the electric load in order to positively impact costs and/or echo-friendliness.
It is a common approach for such systems to regularly sample and store the context of the system in order to back the reasoning algorithms up with historical data.
Figure \ref{linear_time} shows a context model -- represented as a graph -- which is sampled at two different points of time, \textit{t$_{i}$} and \textit{t$_{i+1}$}. 
The graph in each horizontal plane represents the context model at one point of time, where all context variables, independently from their actual values, belong to the same time. 

\begin{figure}[ht]
\centering
\begin{tikzpicture}[scale=0.65]

\begin{axis}[
xlabel=model space,
xlabel style={rotate=-8},
ylabel=model space,
ylabel style={rotate=39},
zlabel=time space, clip=false, xtick=\empty, ytick=\empty, 
ztick={0,1},
zticklabels={t$_{i}$,t$_{i+j}$},
]

\addplot3[mesh, samples=10,domain=0:1]
{x^2*y};

\addplot3+[mesh, scatter, samples=10, mark=*, color=black]
coordinates {
	(0.25,0.5,0.3125) (0.25,0.2,0.0125) (0.5,0.5,0.125) (0.25,0.5,0.3125)
	(0.8,0.5,0.32) (0.5,0.5,0.125) (0.8,0.1,0.064) (0.9,0.7,0.567)    
};

\end{axis}
\end{tikzpicture}
\vspace{-1.0em}
\caption{Time-distorted Model}
\label{distorted_time}
\end{figure}
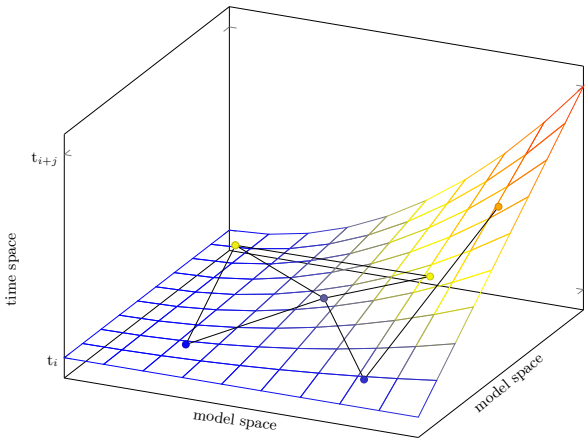

This systematic, regular context sampling, however, has several drawbacks. 
First of all, it yields to a vast amount of data and redundancy, which is very difficult to analyze and process efficiently, especially when correlating the values of different probes sampled at different rates. 
This conflicts with the (near) real-time response time requirements such systems usually need to meet.
Secondly, and more importantly, it is usually not sufficient to simply consider co-temporal data and different values at different moments but should rather be considered simultaneously, {\em e.g.} to investigate a potential causality between two phenomena. 
For instance, predicting the electric load for a particular territory, requires a good understanding of the past electricity production and consumption data in this territory, as well as other data coming from the current context (such as current and forecast weather).
We refer to such context models as time-distorted models, expressing the fact that they contain and union data from different times.
Figure \ref{distorted_time} shows such a time-distorted model.
The context model is again represented as a graph. But this time, the context variables -- again independently from their actual values -- belong to arbitrary different points of time.
This is represented in the figure by a curved, instead of a straight plane.  


Physics, and especially the study of laser~\cite{tom1988time}, relies on a time distortion~\cite{hubral1977time} property, specifying that the observed current time is different depending of the observation context.  
Applied to context models, we claim that the time distortion property could be the fact that a model element, could have different values depending of the observation context, {\em i.e.} depending on the time of the inquiring actor.	
We claim that the time dimension is an important and inherent part of most real-life phenomena and that many reasoning algorithms need to consider the time dimension.  
Further, we claim that the concept of a time-distorted model, which might sound very abstract, is in fact very common and extremely useful, especially for reasoning purposes. 
In order to manage historical context data, we propose to organize context models as a navigable time space, rather than a mere stack of snapshots.	
      		
This paper is organized as follows. 
Section 2 briefly introduces the background of this work and section 3 describes our contribution.
Section 4 presents the evaluation of our approach and the paper concludes, after a discussion of the related work in section 5, with section 6.

%% file: background.tex
\section{Background}

Over the past few years, an emerging paradigm called {\em Models@run.time}~\cite{5280648,5280651} proposes to use models both at design and runtime in order to support intelligent systems.
At design-time, following the Model-Driven Engineering (MDE) paradigm~\cite{DBLP:conf/ifm/Kent02}, models support the design and implementation of the system.
The same (or similar) models are then embedded at runtime in order to support the reasoning processes of intelligent systems, as models offer a {\em simpler, safer and cheaper}~\cite{Rothenberg89thenature} means to reason.
In the remainder of this section, we will integrate the time distortion theory into our Kevoree Modeling Framework~\cite{conf/models/FouquetNMDBPJ12} (KMF~\footnote{http://kevoree.org/kmf}), which is the modeling pillar supporting Kevoree Models@run.time platform~\cite{fouquet:hal-00688707}.

In KMF, each model element can be accessed within the model by a path (form the root element of a model along its containment~\cite{omg2011mof} references to a specific element), which defines the semantic to efficiently navigate in the data space.
The path of a model element may vary with time, typically if it is moved into another container~\cite{omg2011mof}, for example if a component is moved from one node to another one.

Our contribution extends the structural path of model elements with temporal data in order to provide seamless navigation, not only in space but as well in the time space.

%% file: contribution.tex
\section{Contribution}

\subsection{General Idea}

The general goal of our approach is to add the time dimension as a crosscutting concern of data modeling.
Indeed, model elements can independently evolve in time at different paces, so there is no need to enforce the sampling of all model elements at the same rate.
Our solution does not store snapshots of an entire model but rather only stores updated model elements together with their relative time. 
Our hypothesis is that time distortion is part of the domain knowledge itself ({\em e.g.} for wave propagation prediction, electric load prediction) and that therefore, defining and navigating this time distortion directly within domain models is far more efficient and convenient than independently querying each model element with the right time.
Therefore, we provide a natural and seamless navigation concept to navigate into the space-time continuum of data. 
Most importantly, we enable a model to combine elements from different points of time, forming a time-distorted model, which is especially useful for time related reasoning.   
We claim in contrary to classic data mining, our approach can fit with most of (near) real time reasoning requirements.

\subsection{Time Relativity for Model Elements and Relationships}

Our approach is based on data description concepts defined in the MOF~\cite{omg2011mof} standard for model-driven engineering.
We rely on two properties to integrate the time dimension as a crosscutting concern into models: {\em i}) each model element must be uniquely identifiable, and {\em ii}) it must be possible to get a serialized representation of all attributes and relationships of each model element.   

To ensure the first property, KMF defines a \textit{path} for each model element, starting from the root element of a model through its containment relationships to the element, as a unique identifier.
Since the containment graph is actually a tree (each element, except the root, has to be contained exactly once), the path act as a unique full qualified name/ID.
Using this path as an identifier ---instead of a simple string or number--- in particular enables the efficient look-up of model elements.
For our approach, however, it is just important that every model element can be uniquely identified.
This property basically defines the data space continuum and does not include any time information. 
 
\begin{figure}[ht]	
			\centering			
			\includegraphics[scale=0.4]{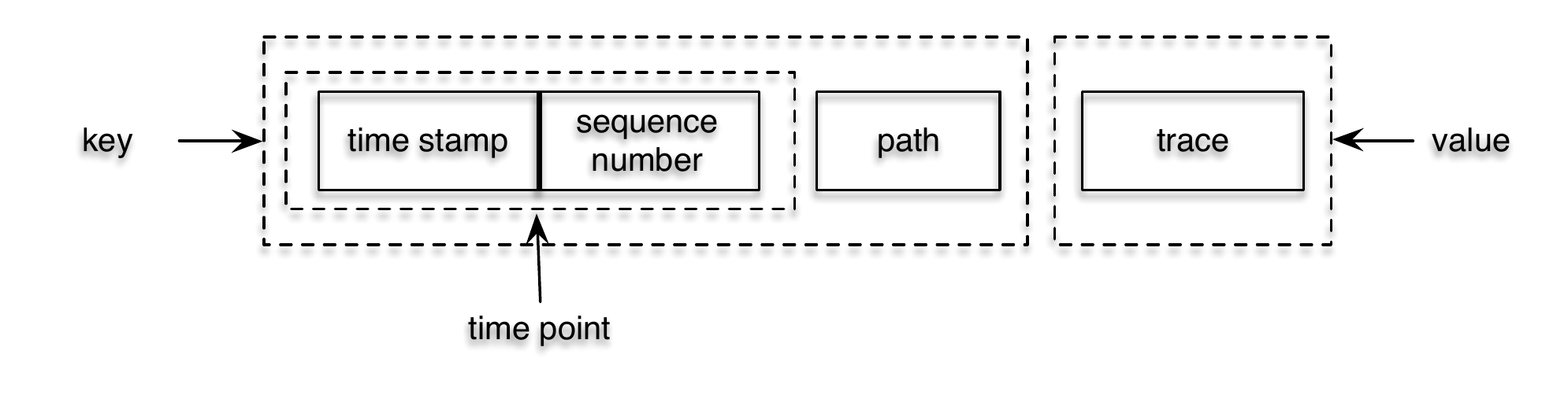}  
			\vspace{-1.5em}
			\caption{Key/value for Time-relative Storage}
			\label{time_aware_principle}   
		\end{figure}

For the second property, KMF serializes models and model elements into so called \textit{traces}. 
A trace defines the sequence of atomic actions in order to construct a model element, using the path concept (including the relationship information).
Each model element can be transformed into a trace and vice versa. 
In addition to that we introduce a time point, which consists of a time stamp and a sequence number. The latter is needed to enable several versions of the same model element at one time stamp. 
This allows us to store and retrieve model elements within their time dimension in a simple key/value format, which can be stored in different back ends, {\em e.g.} key/value stores, relational databases, or simply in-memory (as a cache).
This is shown in figure \ref{time_aware_principle}. 

When the model (a directed graph) is traversed, the related model elements are lazily loaded during the model navigation.
This time related resolution, while traversing the object graph, is completely transparent and hidden behind the methods to navigate in the model space, which are injected by the KMF framework.   
Hereby, each model element is resolved with the time of the model element from which the current navigation is started. 
In other words, model elements are, by default, resolved relatively to the time of the navigating model element. 
For example, when navigating from object \textit{A} at time point \textit{t$_{i}$} to object \textit{B}, object \textit{B} is resolved at time point \textit{t$_{i}$} as well. 
In case at time point \textit{t$_{i}$} either object \textit{B} or the relationship does not exist the {\em prior existing version of B} is returned instead.
By always considering model elements in the context of a specific time point, we create a time dimension for model elements. 
This time relative data resolution, is one of the novel concepts of this contribution, indeed unlike in previous approaches, the navigation function is not constant but yields different results depending on the navigation context (i.e. the time).    

\subsection{Navigating into Time}

Based on the hypothesis that time distortion is part of the domain knowledge, we  provide means to enable an efficient navigation into the time dimension.
Therefore, we define four basic operations for model elements. These operations can be called on each model element.
\begin{itemize}[noitemsep,nolistsep]
\item \textbf{shift}: The {\em shift} operation is the main possibility to navigate a model element through the time space. 
It takes a time point as parameter, look for the model version of itself at the required time point, loads the corresponding trace from the storage and replaces the current model element with the one from the storage. 
\item \textbf{deep shift}: The {\em deep shift} operation is similar to the shift operation but not only shifts the model element itself to the required time point but also all its contained model elements.  
\item \textbf{previous}: The {\em previous} operation is a shortcut to retrieve the direct (in terms of time) predecessor of the current model element {\em i.e.}, its previous value.
\item \textbf{next}: The {\em next} method is similar to the {\em previous} operation but retrieves the direct (in terms of time) successor of the current model element. 
\end{itemize}

In addition to these operations, each model element is aware of its relative {\em now} time point. 
This is the time point in which the model element is currently resolved from. 
From the perspective of a model element it is the present time. 
From this model element and this {\em now} point, the navigation context give a list of related reachable elements, from the same or different time.
This distortion in term of navigable relation define what we call a time distorted model.
As result, a move in time operation has just to define an element associated to another time point, and then the time related resolution during data traversing will seamless reach it.
This relative navigation enables a seamless and transparent navigation into the models space-time continuum.

%% file: evaluation.tex
\section{Evaluation}
Our data space-time continuum approach is especially useful for runtime reasoning purposes.
To quantify the advantages of our novel modeling concept, we evaluate it on a smart grid reasoning engine.

\textbf{Smart grids} are emerging infrastructures characterized by the introduction of reactive entities modernizing the legacy electricity distribution grid.
Smart meters, entities installed at customers' sites to continuously measure consumption data and propagate it through network communication links, are one of the main building blocks of smart grids.
The electrical consumption is regularly sampled (\textit{e.g.} every 20 minutes), which leads to a huge context model.
We focus our evaluation on a local over-voltage prediction.
This prediction value can be used to enable/disable devices to smooth the electric consumption at block level (\textit{e.g.} managing around 100 houses).
Due to the propagation time, a context model is inherently composed of different values coming from different points of time.
Our experimental validation focuses on the performance (execution time) of a reasoning process which regularly analyzes the last 20 measured values and determines if power consumption has significantly increased.
We summarize our evaluation in two research questions: \textbf{(1)} performance of the insertion time in the data space-time continuum? \textbf{(2)} performance of the reasoning process, especially to lookup temporal data?

\textbf{Experimental results:} we implemented the reasoning engine case study using KMF. 
It has been implemented twice, once with a full model sampling, and once with our space-time continuum strategy using a time-distorted model. 
The reasoning context model under study contains 100 smart meters with 10000 values history each, resulting in 1 million elements to store and analyze. 
Note that data are stored to disk at each sampling.
The execution time on a MacBook Pro i5 2.4Ghz, 16GB RAM, is presented bellow:
\vspace{2mm}

\scriptsize
\begin{tabular}{|l|r|r|}
  \hline
   \textbf{Modeling strategy} & \textbf{Insert time} & \textbf{Reasoning time} \\
  \hline
  Sampling & 265.9 s & 162 ms \\
  Space-time continuum & 16.1 s & 4 ms\\
  \hline
\end{tabular}
\normalsize
\vspace{2mm}

As shown in the table, our space-time continuum strategy leads to a reduction of the reasoning time by a factor of \textbf{40}, compared to the classic sampling strategy. 
The insert time (for storing the context values) has also been significantly improved by a factor of \textbf{16}.
Our experimental smart grid reasoning engine can thus meet most of (near) real time requirements.

%% file: relatedwork.tex
\section{Related Work}

The lack of a temporal dimension in data modeling has been discussed in details, especially in the area of databases. 
In an early work Clifford {\em et al.}~\cite{conf/xp/CliffordW81} provide a formal semantic for historical databases. 
They present an intentional logic as a formalism for expressing the temporal semantics.
Rose and Segev~\cite{conf/er/RoseS91} suggest to extend the entity-relationship data model into a temporal, object-oriented one, incorporating temporal structures and constraints in the data model itself rather than at the application level. They also propose a temporal query language for the model. 
Ariav~\cite{Ariav1986Temporally} suggests a temporally-oriented data model that is a restricted superset of the relational model. 
He adds a temporal aspect to the tabular notion of data and provides a framework and a SQL-like query language for storing and retrieving data, taking their temporal context into account.
The works of Mahmood {\em et al.}~\cite{journals/corr/abs-1002-1143} and Segev and Shoshani~\cite{conf/ssdbm/SegevS88} take a similar direction and seek to extend the relational model with temporal aspects.  
In an earlier work Segev and Shoshani~\cite{Segev:1987:LMT:38713.38760} examine the semantics of temporal data and corresponding operators independently of any traditional data model.
Similar to these works, our approach proposes to add the time dimension into data modeling.
In addition to this, we not only store (and query) historical data, but we also propose a way to use time-distorted data sets for intelligent reasoning.  
Also, we do not extend an existing data model ({\em e.g.} the relational data model) with temporal structures but use model-driven engineering techniques to integrate the time dimension as a crosscutting property of any model element.
Furthermore, we do not rely on a complex query language for retrieving temporal data.
Instead, our approach aims at providing a more natural, query-less and seamless navigation into the time dimension of each model element, allowing a composition of different time-related values to build a dedicated context model for reasoning purposes (inspired by temporal logic~\cite{4567924}).    
Like version control systems, {\em e.g.} Git~\cite{git}, our approach only stores incremental changes (over time) rather than snapshots of the complete system. 

%% file: conclusion.tex
\section{Conclusion}

Considering time as a crosscutting concern of data has been discussed since more than two decades ~\cite{Ariav1986Temporally, conf/er/RoseS91}, especially in the area of databases~\cite{journals/corr/abs-1002-1143, conf/ssdbm/SegevS88}. 
However, recent data modeling approaches mostly rely on a discrete time representation, which can hardly consider model elements ({\em e.g.} context variables) coming from different points of time. 
In this paper, we presented a novel modeling approach which considers time as a first-class property crosscutting any model element, and which makes it possible to organize a model as a data space-time continuum. 
We also presented a time-distorted context model, which enables the combination of elements from different points of time, and thus form a dedicated time-distorted view of the data continuum.
Instead of introducing a dedicated querying language, as for example proposed by ~\cite{conf/er/RoseS91,Ariav1986Temporally}, we provide a seamless and transparent navigation mechanism. 
In particular, our novel modeling technique leverages a time-relative navigation between model elements in order to fill the gap between classic data modeling and time consideration.
Our approach has been implemented and integrated into an open source modeling framework, named KMF, and evaluated with a smart grid reasoning engine for electric load prediction. 
We showed that our approach supports time-based reasoning processes and allows them to be compatible with most of (near) real-time requirements.
For future work, we plan to study domain specific time distortion patterns, like wave propagation prediction, and evaluate how we can leverage our approach to support reasoning processes for such phenomenons. 